\documentstyle[twoside,fleqn,espcrc2,epsfig,pictex]{article}

\newcommand{\AmS}{{\protect\the\textfont2
  A\kern-.1667em\lower.5ex\hbox{M}\kern-.125emS}}
\newcommand{\D}{D\kern-2.2mm/}

\title{Very Light Axion}

\author{Jihn E. Kim\address{Department of Physics and Center for
Theoretical Physics, Seoul National University,\\ 
Seoul 151-742, Korea \ (e-mail address: {\it jekim@phyp.snu.ac.kr})}
\address{School of Physics, Korea Institute for
Advanced Study,\\ 
207-43 Cheongryangri-dong, Seoul 130-012, Korea}
        \thanks{This work is supported by Distinguished Scholar
Exchange Program of Korea Research Foundation, KOSEF and Ministry of 
Education
BSRI 97-2468.}
        }
\begin{document}       

\begin{abstract}
I review the idea of axion, axion properties, and
the superstring axion.
\end{abstract}

% typeset front matter (including abstract)
\maketitle

\section{INTRODUCTION}

Quantum chromodynamics (QCD) before 1975 was described by
\begin{equation}
{\cal L}=-{1\over 2g^2}{\rm Tr} F_{\mu\nu} F^{\mu\nu}
+\bar q(i\D -M)q
\end{equation} 
But after 1975, due to the discovery of instanton solution in
non-abelian gauge theories \cite{belavin}, it has been known that 
one should consider another term also\cite{callan}  
\begin{equation}
{{\bar \theta}\over 16\pi^2}{\rm Tr}F_{\mu\nu}\tilde F^{\mu\nu}
\end{equation}
which is ${\bf E}^a\cdot{\bf B}^a$ and violates the CP invariance
in strong interactions. $|\bar\theta|$ is phenomenologically
bounded, by the upper bound of the neutron electric dipole
moment $|d_n|<10^{-25}e$cm,
\begin{equation}
|\bar\theta|<10^{-9}
\end{equation}
Here, the question arises, $\lq\lq$Why is $|\bar\theta|$ so small?",
which is the parameter problem in the standard model. It is
commonly called {\it the strong CP problem}.
Most small parameters in physics have led to some ideas, mostly those
related to symmetries. For example, the small ratio $M_W/M_P\ll 1$
(the so-called gauge hierarchy problem) led to supersymmetric
solution \cite{weinberg}, and $m_u,m_d\ll 1$ GeV led to the
chiral symmetry \cite{gor}. 

The nicest solution of the strong CP
problem is the existence of {\it axion} (very light, or invisible(?)
axion). The axion is related to the Pecei-Quinn symmetry \cite{pq}.
The axion relevant for the standard model VEV is the PQWW axion
\cite{pqww}. Since it has not been observed \cite{obs}, many calculable
$\bar\theta$ models were tried in 1978 \cite{cal}. Because the ideas
without axion were turned out to be ugly, one freed the 
axion scale from the electroweak scale, inventing a very light 
(or invisible) axion \cite{inv}. ¤
For a very light axion, the model-building is one of the
important aspects, since the axion decay constant is allowed
within the window from SN1987A constraint and closure density
of the universe \cite{this},
\begin{equation}
10^{9-10}\ {\rm GeV}\le F_a\le 10^{12-13}\ {\rm GeV}
\end{equation}

In this talk, I review why axion solves the strong CP problem,
with a comment on the very light axion in superstring models.

\section{THE AXION SOLUTION}

Peccei and Quinn introduced an anomalous chiral symmetry \cite{pq},
$U(1)_A$. If this global symmetry is spontaneously broken,
there results a Goldstone boson, called axion \cite{pqww}. This
axion solution is a dynamical solution of the strong CP problem,
in which the evolving universe settles $\bar\theta=0$. If the
evolving universe settles $\bar\theta$ at 0, $\bar\theta=0$
must be the minimum of the potential. An elegant proof of this
fact has been given by Vafa and Witten \cite{vw} which is
briefly repeated here. After this proof, we show how axion
solves the strong CP problem.

Below the $SU(2)\times U(1)$ symmetry breaking scale, we
have
\begin{equation}
{\cal L}=-{1\over 4}F^2+\bar q(i\D-M_q)q+\bar\theta 
g^2\{F\tilde F\}
\end{equation}
where $\{F\tilde F\}$ means $(1/ 32\pi^2)F^a_{\mu\nu}
\tilde F^{a\mu\nu}$, and $M_q$ is a diagonal, $\gamma_5$ free,
and real quark mass matrix. We now understand that $\bar\theta$
is a dynamical field in axion physics, but for a moment let us
treat it as a parameter. After integrating out the quark 
fields, we obtain the following generating functional
in the Euclidian space
\begin{equation}
\int [dA_\mu]\prod_i {\rm Det}(\D+m_i)e^{-\int
d^4x({1\over 4g^2}F^2-i\bar\theta\{F\tilde F\})}
\end{equation}
The Euclidian Dirac operator satisfies that if 
$i\D \psi=\lambda\psi$ then $i\D(\gamma_5\psi)=-\lambda
(\gamma_5\psi)$. Namely, if $\lambda$ is a real eigenvalue of
$i\D$ then so is $-\lambda$, which implies 
\begin{eqnarray}
&{\rm Det}(\D+m_i)=\prod_{\lambda}(i\lambda+m_i)\nonumber\\
&=m_i^{N_0}\prod_{\lambda>0}(m_i^2+\lambda^2)>0
\end{eqnarray}
where $N_0$ is the number of zero modes. Therefore, we obtain
the following inequality using the Schwarz inequality,
\begin{eqnarray}
&e^{-\int d^4xV[\bar\theta]}\equiv
 \Big|\int [dA_\mu]\prod_{i}{\rm Det}(\D+m_i)\cdot\nonumber\\
&\cdot e^{-\int d^4x {\cal L}(\bar\theta)}\Big|\nonumber\\
&\le \left|\int [dA_\mu]\prod_{i}{\rm Det}(\D+m_i)
e^{-\int d^4x{\cal L}(\bar\theta=0)}\right|\\
&=e^{-\int d^4xV[0]}\nonumber
\end{eqnarray}
which implies that
\begin{equation}
V[\bar\theta]\ge V[0]
\end{equation}
The schematic form of $V[\bar\theta]$ is shown in Fig. 1, where
the height of $V$ is $(2Z/(1+Z)^2)f_\pi^2m_\pi^2$ with $Z=m_u/m_d$.
Note that it is a periodic function of $\bar\theta$ with period
$2\pi$ since $Z\propto \int [dA_\mu](\cdots)e^{i\bar\theta
\int d^4x\{F\tilde F \}}$ and $\int d^4x \{F\tilde F\}={\bf Z}$.

%
% FIGURE 1 
%
\begin{figure}[htb]
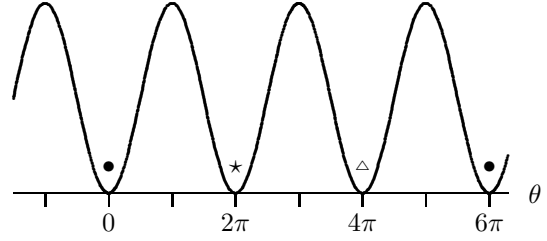

%\vspace{9pt}
%\framebox[55mm]{\rule[-21mm]{0mm}{43mm}}
$$\beginpicture
\setcoordinatesystem units <72pt,36pt> point at 0 0
\setplotarea x from -0.500 to 2.100, y from 0.000 to 2.000
\axis bottom %label {$\theta$}
      ticks out
      width <0.5pt> length <5.0pt>
      withvalues {} {$0$} {} {$2\pi$} {} {$4\pi$} {} {$6\pi$} {} /
      at -0.333 0.000 0.333 0.666 1.000 1.333 1.666 2.000 /
/
\put {$\theta$} <10pt,0pt> at 2.100 0.000
\setplotsymbol ({\normalsize.})
\setsolid
\setquadratic
\plot
-0.500  1.000 -0.450  1.454 -0.400  1.809 -0.350  1.988 -0.300  1.951
-0.250  1.707 -0.200  1.309 -0.150  0.844 -0.100  0.412 -0.050  0.109
 0.000  0.000  0.050  0.109  0.100  0.412  0.150  0.844  0.200  1.309
 0.250  1.707  0.300  1.951  0.350  1.988  0.400  1.809  0.450  1.454
 0.500  1.000  0.550  0.546  0.600  0.191  0.650  0.012  0.700  0.049
 0.750  0.293  0.800  0.691  0.850  1.156  0.900  1.588  0.950  1.891
 1.000  2.000  1.050  1.891  1.100  1.588  1.150  1.156  1.200  0.691
 1.250  0.293  1.300  0.049  1.350  0.012  1.400  0.191  1.450  0.546
 1.500  1.000  1.550  1.454  1.600  1.809  1.650  1.988  1.700  1.951
 1.750  1.707  1.800  1.309  1.850  0.844  1.900  0.412  1.950  0.109
 2.000  0.000  2.050  0.109  2.100  0.412
/
\put {$\bullet$} <0pt,10pt> at 0.000 0.000
\put {$\star$} <0pt,10pt> at 0.667 0.000
\put {\tiny$\triangle$} <0pt,10pt> at 1.333 0.000
\put {$\bullet$} <0pt,10pt> at 2.000 0.000
\endpicture$$
\caption{An axionic potential with $N_{DW}=3$. Here $\theta=a/F_a$.}
\label{fig.1}
\end{figure}
%\vskip 0.3cm
 
At this stage, any $\bar\theta$ will be a good theory, i.e.
any nonzero value of $\bar\theta$ is allowed.

The axion solution is to identify $\bar\theta$ as a dynamical
field. Interpreting it as a spin zero boson field $a$,
\begin{equation}
\bar\theta\equiv {a\over F_a}
\end{equation}
one should introduce a scale $F_a$ which is called {\it the axion
decay constant}. In this axion theory, the evolving universe
allow only one stable value of $\bar\theta$, i.e. the minimum
value of $V$ at $\bar\theta=0$. In this case, different $\bar\theta$'s
do not describe different theories, but only different vacua.
An important feature of the above proof is that {\it $\bar\theta$
does NOT have any potential except that coming from $\bar\theta\{
F\tilde F\}$.} Otherwise, the mechanism does not work.

To make $\bar\theta$ dynamical, one must have a mechanism to
introduce a kinetic energy term for $\bar\theta$ and the scale
$F_a$. Depending on this origin, one can classify axions as
: (i) Goldstone boson of a spontaneously broken anomalous
$U(1)_A$ ($F_a$ is the spontaneous symmetry breaking scale), 
(ii) fundamental field in string theories ($F_a$ is the 
compactification scale), and (iii) composite field ($F_a$ is
the confinement scale).

The most plausible model is the superstring axion, which will
be briefly commented later. 
  
Another important parameter of the axion model is the domain
wall number $N_{DW}$. It arises because in the field space
all fields return to the original value after the chiral
rotation (of $U(1)_A$ group) of $2\pi N_{DW}$. For example,
the vacua $<a>=0, 2\pi F_a, 4\pi F_a,$ etc gives the minima of $V$, but
$<a>=0$ is identified only at $2\pi N_{DW}F_a\times$(integer).  

\section{AXION PROPERTIES}

Remembering that the axion is a dynamical $\theta$ \footnote{From 
now on we suppress $\ \bar {}\ $ in $\bar\theta$}, we
can easily derive its mass and interaction terms. Here,
we derive the mass term only in effective field theory framework.

The simplest axion model is the heavy quark axion model since
at low energy it introduces only the dynamical $\bar\theta$
in addtition to the standard model fields. The Lagrangian is
\begin{equation}
{\cal L}=\sigma \bar Q_RQ_L+h.c.+\cdots
\end{equation}
where we suppressed the Yukawa coupling constant and $\cdots$
are the other terms, including the potential respecting the
$U(1)_A$ symmetry. The PQ global $U(1)_A$ symmetry is
\begin{eqnarray}
&Q_L\rightarrow e^{-i{\alpha\over 2}}Q_L,\ Q_R\rightarrow
e^{i{\alpha\over 2}}Q_R,\ \sigma\rightarrow e^{i\alpha}
\sigma\nonumber\\
&\theta\rightarrow\theta-\alpha
\end{eqnarray}
For a nonzero VEV $<\sigma>=F_a/\sqrt{2}$, $Q$ obtains a mass
at scale $F_a$, the radial component $\rho$ (Higgs--type field) 
of
$\sigma$ obtains a mass at scale $F_a$, and at low energy 
there remains only the axion $a$. Thus from the kinetic 
term $D_\mu\sigma^*D^\mu\sigma$, we obtain $(1/2)\partial_\mu
a\partial^\mu a$ where $\sigma=([F_a+\rho]/\sqrt{2})e^{ia/F_a}$.

Thus, below the scale $F_a$, the light fields are gluons and
$a$ (plus the other SM fields). The relevant part of the
Lagrangian respecting the symmetry (12) (i.e. with $a
\rightarrow a+\alpha F_a$) is
\begin{eqnarray}
&{\cal L}={1\over 2}(\partial_\mu a)^2+{\rm (derivative\ term\ 
of\ } a)\nonumber\\
 &+(\theta+{a\over F_a})\{F\tilde F\}
\end{eqnarray} 
Note that we created the needed 
$F\tilde F$ coupling minimally. Usually,
$a$ is redefined as $a-\theta F_a$ so that the coefficient of
$F\tilde F$ is $a/F_a\equiv\bar\theta$.

\subsection{Axion Mass in One Flavor QCD}

To see the essence of the mass formula, let us consider the
one-flavor QCD,
\begin{equation}
{\cal L}=-m_u\bar u_Ru_L+{\rm h.c.}
\end{equation}
which possess the following hypothetical symmetry
\begin{eqnarray}
&u_L\rightarrow e^{i\alpha}u_L,\ \bar u_R\rightarrow e^{i\alpha}
\bar u_R,\ m\rightarrow e^{-2i\alpha}m\nonumber\\
&\theta\rightarrow \theta+2\alpha 
\end{eqnarray}
Since $m$ is endowed with a transformation even
though it is not a symmetry, Eq. (15) is
useful to trace the $m$ dependence in the effective theory
below the quark condensation scale $<\bar uu>\propto v^3e^{i\eta/v}$,
\begin{eqnarray}
&V={1\over 2}m_u\Lambda^3e^{i\theta}-{1\over 2}\lambda_1
\Lambda v^3e^{i(\eta/v-\theta)}\nonumber\\
&-{1\over 2}\lambda_2m_uv^3e^{i\eta/v}
+\lambda_3m_u^2\Lambda^2e^{2i\theta}\\
&+\lambda_4{v^6\over\Lambda^2}
e^{2i(\eta/v-\theta)}+\cdots+{\rm h.c.}\nonumber
\end{eqnarray}
where the strong interaction scale $\Lambda$ is inserted to make
the dimension appropriate.
Note that, for $m_u=0$, $\eta-F_a\theta$ can be redefined as a new
$\eta$, removing the $\theta$ dependence. Thus $\theta$ is
unphysical in a massless $u$-quark theory, solving the strong
CP problem. At the minimum of the potential, the $a$--$\eta$
mass matrix for $m_u\ne 0$ is
\begin{equation}¤
M^2=\left(\matrix{&\lambda\Lambda v+\lambda' mv, 
&-\lambda\Lambda v^2/F_a\cr
&-\lambda\Lambda v^2/F_a, 
&-{m\Lambda^3\over F_a^2}+{\lambda\Lambda v^3\over F_a^2}\cr}\right)
\end{equation}
Diagonalizing the above mass matrix for $F_a\gg$ (other mass
parameters), we obtain for vacuum at $\theta=0$
\begin{eqnarray}
&m_a^2={m_u\Lambda\over F_a^2}\left({\lambda\lambda' v^4\over
\lambda\Lambda v+\lambda' m_uv}-\Lambda^2\right),\nonumber \\ 
&m_\eta^2=(\lambda\Lambda+\lambda' m)v
\end{eqnarray}
which shows the essential features of the axion mass: it is
suppressed by $F_a$, multiplied by $m_q$, and the rest of
condensation parameters. If the above mass squared is negative,
we chose a wrong vacuum and choose $\theta=\pi$ 
instead as the vacuum.

\subsection{Axion Mass in Two-flavor QCD}

For a realistic axion mass, we consider one family QCD, in which
we assign $U(1)_u\times U(1)_d$ fictitious symmetry,
\begin{eqnarray}
&u_L\rightarrow e^{i\alpha}u_L,\ d_L\rightarrow e^{i\beta}d_L,
\ m_u\rightarrow e^{-2i\alpha}m_u,\nonumber\\
&m_d\rightarrow e^{-2i\beta}m_d,\ \theta\rightarrow \theta
+2(\alpha+\beta)
\end{eqnarray}
Following the same procedure as in the previous subsection, we
obtain
\begin{eqnarray}
&V=m_um_d\Lambda^2\cos{a\over F_a}-\lambda_1{v^6\over\Lambda^2}
\cos\left({2\eta\over F_\pi}-{a\over F_a}\right)\nonumber\\
&-\lambda_2 m_uF_\pi^3\cos\left({\eta\over F_\pi}
+{\pi_0\over F_\pi}\right)\nonumber\\
&-\lambda_3m_dF_\pi^3\cos
\left({\eta\over F_\pi}-{\pi^0\over F_\pi}\right)\\
&-\lambda_4m_uF_\pi^3\cos\left({\eta\over F_\pi}
-{\pi^0\over F_\pi}-{a\over F_a}\right)\nonumber\\
&-\lambda_5m_dF_\pi^3\cos\left({\eta\over F_\pi}
+{\pi^0\over F_\pi}-{a\over F_a}\right)+\cdots\nonumber
\end{eqnarray}
Separating two mass eigenstates, we obtain the usual
axion mass formula \cite{bt}
\begin{equation}
m_a={m_{\pi^0}F_\pi\over F_a}{\sqrt{Z}\over 1+Z}
\end{equation}
where $Z=m_u/m_d$. The above formula is valid for the
KSVZ model. For the PQWW and DFSZ models,
one needs extra consideration, for removing the 
longitudinal component of $Z^0$. In the limit of
$F_a\gg$ (other mass parameters), i.e. in the DFSZ model,  
the above formula is also valid.

\subsection{Models}

\vskip 0.3cm
\noindent{\bf KSVZ Model}

This simplest axion model was already introduced in Eqs. (11)--(13).
Here we write the $U(1)_A$ current resulting from that
Lagrangian,
\begin{eqnarray}
J_\mu \ & =
\ \  \tilde v\partial_\mu a-{1\over 2}\bar Q\gamma_\mu\gamma_5Q 
\nonumber\\
&+{1\over 2(1+Z)}(\bar u\gamma_\mu\gamma_5u+Z\bar d\gamma_\mu\gamma_5d)
\end{eqnarray}
Note that the last term results from the $a,\pi^0,\eta$ diagonalization
procedure. There is no $a$--electron--electron coupling in this
model, but can arise at loop orders.

\vskip 0.3cm
\noindent{\bf DFSZ Model}

Here one uses a singlet scalar $\sigma$, and two Higgs doublets,
\begin{eqnarray}
&{\cal L}=\sigma\sigma H_1^{0*}H_2^{0*}-\bar u_Ru_LH_2^{0*}\nonumber\\
&+\bar d_Rd_LH_1^{0*}+\bar e_Re_LH_1^{0*}+\cdots
\end{eqnarray}
where couplings are suppressed. This model has another
parameter, i.e. the ratio of VEV's of Higgs doublets,
\begin{equation}
x={v_2\over v_1}=\tan\beta
\end{equation} 
This is the so-called 
$(d^c,e)$--unification model in which the electron obtains mass
through the Higgs doublet giving mass to the $d$--quark. One
can similarly define a $(u^c,e)$--unification model, and a
non-unification model (a third Higgs doublet gives mass to electron).
In this model, the axion mostly resides in $\sigma$, but $H_1^0$
and $H_2^0$ also contain small component of $a$. Thus, 
$\bar ei\gamma_5 ea$ coupling arises at tree level,
\begin{equation}
{a\over F_a}m_e{2x\over x+x^{-1}}\bar ei\gamma_5 e
\end{equation}
The axion current is given by
\begin{eqnarray}
J_\mu &=\tilde v\partial_\mu a+{x^{-1}\over x+x^{-1}}\sum_{i}
\bar u_i\gamma_\mu\gamma_5u_i\nonumber\\
&+{x\over x+x^{-1}}\sum_{i}\bar d_i\gamma_\mu\gamma_5d_i+(\cdots)
\end{eqnarray}
where $(\cdots)$ is the last term in Eq. (22), arising in the
process of $<q\bar q>$ condensation.

Detection through cavity experiments is the relevant phenomenology for
very light axions.

\vskip 0.3cm
\noindent {\bf PQWW Axion}

For the PQWW axion, $v_1=\sqrt{2}<H_1^0>$ and $v_2=
\sqrt{2}<H_2^0>$ break both $U(1)_Y$ gauge and
$U(1)_A$ global symmetries. Following the similar, but
more complicated algebra, one obtains
\begin{equation}
F_a={\sqrt{v_1^2+v_2^2}\over (x+x^{-1})N_g}
\end{equation}
where $N_g$ is the family number. For $x=1$ and $N_g=3$,
$m_a\sim 150$ keV; so it does not decay to $e^+e^-$ pair.
For other values of $x$, the axion mass is greater. For
example, for $x>30$ (which is one possible region of parameter
space in MSSM model) $m_a>$ 2.3 MeV. But current data
prefer $x$ near 2.

Particle physics phenomenology excludes the PQWW axion. Early
examples are the reactor experiments, $K^+$ decay, $\pi^+$
decay, etc. \cite{peccei}. The argument was 
made more concrete later \cite{pec1}.
Note, however, that the astrophysical bound $m_a<10$ meV ($F_a>
10^9$ GeV) is better than these particle physics arguments.
$K^+\rightarrow \pi^+a$ gives $F_a>10^{3-4}$ GeV. In $J/\psi$,
$\Upsilon$ decays, we consider the ratio
\begin{eqnarray}
&R_Q={\Gamma(Q\bar Q)\rightarrow a\gamma\over\Gamma(Q\bar Q)
\rightarrow\mu^+\mu^-}={G_Fm_Q^2\over\sqrt{2}\pi\alpha}\cdot\nonumber\\
&\cdot\left[1-{({\pi^2\over 2}+8\ln 2)\alpha_s\over 3\pi}\right]Z^2_Q
\nonumber
\end{eqnarray} 
where $Z_Q=x,x^{-1}$ for $Q=2/3,$ and --1/3 quarks, respectively.
Then B.R.($\Upsilon\rightarrow a\gamma$)$=(2.0\pm 0.7)\times 10^{-4}
Z_b^2$, and B.R.$(\psi\rightarrow a\gamma)=(3.7\pm 0.8)\times 10^{-5}
Z_c^2.$ On the other hand, the experimental upper bounds are
B.R.($\Upsilon\rightarrow a\gamma$)$<3\times 10^{-4}$ and
B.R.($\psi\rightarrow a\gamma$)$<1.4\times 10^{-5}$. 
Therefore, $x\sim x^{-1}$ from the above condition. 
The variant axion models \cite{var} can have $Z_b,c\propto x,x^{-1}$,
leading to a large axion mass. For $m_a>3m_e$, the decay
$\pi^+\rightarrow e^+e^-e^+\nu_e$ can be used to rule out
short-lived $\lq\lq$visible" axion models \cite{pec1}. 

\subsection{$a\gamma\gamma$ Coupling}

For the very light axion models, the axion--photon--photon
coupling is represented by two pieces \cite{kaplan}:
$\bar c_{a\gamma\gamma}$ which is defined by the
global symmetry of the Lagrangian (i.e. term given by the
short distance scale) and the piece added below the quark
condensation scale (i.e. the term added in the effective
theory at long distance scale),
\begin{eqnarray}
c_{a\gamma\gamma}=&\bar c_{a\gamma\gamma}-{2\over 3}{4+Z\over
1+Z}\nonumber\\
&=\bar c_{a\gamma\gamma}-1.92
\end{eqnarray}
where we used $Z=0.6$.  Since $\bar c_{a\gamma\gamma}$ is 
extracted from the coefficient of 
$(a/F_a)\{F_{e.m.}\tilde F_{e.m.}\}$, one can easily convince
himself that
\begin{equation}
\bar c_{a\gamma\gamma}={E\over C}
\end{equation}
where $E={\rm Tr}Q_{em}^2Q_{PQ}$ and $\delta_{ab}C=
{\rm Tr}\lambda_a\lambda_b Q_{PQ}$, and $Q_{PQ}$ is
the global $U(1)_A$ charge given by the Lagrangian. For a
fundamental representation of $SU(N)$, we define the
index $\ell=1/2$. Thus we obtain
\begin{eqnarray}
{\bf KSVZ:}\nonumber\\
&\ C_3=-{1\over 2}, C_8=-3,\nonumber\\
&\ E_3=-3e_Q^2,E_8=-8e_Q^2\nonumber\\
&\ \bar c_{a\gamma\gamma}=\left\{\matrix{&6e_Q^2\ &{\rm for}\ {\bf 3,3^*}\cr
&{8\over 3}e_Q^2\ &{\rm for}\ {\bf 8}}\right.\nonumber\\
\nonumber\\
{\bf DFSZ:}\nonumber\\
&\ C_3=N_g, E={8\over 3}N_g\ \ \ {\rm for}\ (d^c,e)
\nonumber
\end{eqnarray}
\vskip 0.3cm
Here $C_x$ and $E_x$ are calculated for the representation {\bf x}
and $e_x$ in the KSVZ is the electromagnetic charge in
units of positron charge of the heavy
quark $Q$ transforming as {\bf x} of $SU(3)_c$. $c_a\gamma\gamma$
is given in Table 1 for several interesting cases \cite{kim98}.

\begin{table*}[hbt]
% space before first and after last column: 1.5pc
% space between columns: 3.0pc (twice the above)
\setlength{\tabcolsep}{1.5pc}
% -----------------------------------------------------
% adapted from TeX book, p. 241
\newlength{\digitwidth} \settowidth{\digitwidth}{\rm 0}
\catcode`?=\active \def?{\kern\digitwidth}
% -----------------------------------------------------
\caption{$c_{a\gamma\gamma}$ in KSVZ and DFSZ models.
The unification condition in DFSZ is which Higgs doublet 
couples to $e$.}
\label{tab:effluents}
\begin{tabular*}{\textwidth}{@{}l@{\extracolsep{\fill}}rrrr}
\hline
                 & \multicolumn{2}{l}{KSVZ models} 
                 & \multicolumn{2}{l}{DFSZ models} \\
\cline{2-3} \cline{4-5}
                 & \multicolumn{1}{r}{Heavy quark charge(s)} 
                 & \multicolumn{1}{r}{$c_{a\gamma\gamma}$} 
                 & \multicolumn{1}{r}{($x=<H_2^0>/<H_1^0>$)} 
                 & \multicolumn{1}{r}{$c_{a\gamma\gamma}$}         \\
\hline
 & $ e_Q=0$ & $-1.92$ & $ (d^c,e)\ unif.\ ({\rm any}\ x) $ & $0.75$ \\
 & $ e_3=-1/3$ & $-1.25$ & $(u^c,e)\ unif.\ (x=1)$ & $-2.17$ \\
 & $ e_3=2/3$ & $0.75$ & $(u^c,e)\ unif.\ (x=1.5) $ & $-2.56$ \\
 & $ e_3=1$ & $4.08$ & $(u^c,e)\ unif.\ (x=60)$ & $-3.17$ \\
 & $ e_8=1$ & $0.75$ & $nonunif.\ (x=1)  $ & $-0.25$ \\
 & $ (e_3=2/3)+(e_3=-1/3)$ & $-0.25$ & $nonunif.\ (x=1.5)$ & $-0.64$ \\
                 & & & $nonunif.\ (x=60)$ & $-1.25$ \\ 
\hline
\end{tabular*}
\end{table*}

In reality such as in superstring models, there can be 
many heavy quarks which carry nontrivial PQ charges. 
For example, superstring models have $\sim$ 400 chiral
fermions. Here, light fields consist of the usual
45 chiral fields, two Higgs doublets ($H_1, H_2$), 12
gauge bosons, and their superpartners. If $H_1$ and
$H_2$ carry the PQ charges, then there is the aspect of
DFSZ. On the other hand, the heavy fields consist of
$\sigma, (3,1,1)+(3^*,1,1), (1,2,1)+(1,2,1)$, etc. Among
these heavy fields, $\sigma$ is present. If some
color triplets and antitriplets carry the PQ charges,
then there exists the KSVZ aspect. Therefore, 
most probably both KSVZ and DFSZ effects
add up and one should calculate $c_{a\gamma\gamma}$
for a specific model. Let us hope that there will
appear in the future a {\it standard superstring model} where one
can predict $c_{a\gamma\gamma}$ unambiguously.

%
% Fig. 2 
%
\begin{figure}[htb]
%\vspace{9pt}
%\framebox[55mm]{\rule[-21mm]{0mm}{43mm}}
\centerline{\epsfig{file=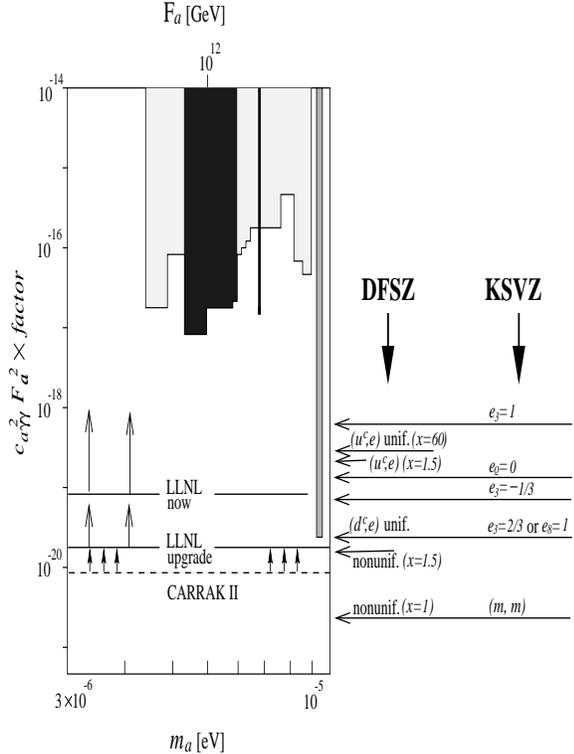,width=75mm,height=100mm}}
\caption{Data and KSVZ and DFSZ model predictions}
\label{fig.3}
\end{figure}

In Fig. 2, we compare these predictions of very light axion 
models and cosmic axion detection experiments
of RBF \cite{rbf}, UF \cite{uf}, LLNL \cite{llnl},
and Kyoto \cite{kyoto} with several model predictions. 
Fig. 3 is  the data collection.
Note that we are at 
the edge of finding the very light axion. One caviat in this
comparison is that one uses the cosmological
estimation of the axion density around earth \cite{turner}.

\section{OTHER SOLUTIONS}

The strong CP problem is to understand the reason why $\theta$
is so small.  There is another method to make $\theta$ small.
Firstly, start ${\cal L}$ with vanishing $\theta$. It can be done in
models with CP invariant Lagrangian. Second, introduce CP violation
in weak interactions. It must be done by spontaneous CP violation
\cite{lee}. Then, at higher orders $\theta$ will be generated,
which must satisfy the experimental bound. One loop $\theta$
is typically of order $(10^{-4}-10^{-5})\times$(couplings) 
which is usually too large. It is
therefore desirable to forbid $\theta$ even at one loopi
\cite{cal}. Of course,
the tree level contribution must be absent, which is implemented
by making Arg.Det.$M_q=0$. Then calculate $\theta$ at one loop
and convince oneself that the final $\theta$ is sufficiently
small. Therefore, these are called calculable models. 

The PQWW axion was looked for in 1978 and it seemed had troubles.
Because of the trouble, the calculable models were studied
extensively in 1978.
But this calculable models are not as attractive as the automatic
solution based on axion. Thus in 1979 the very light axion model
was invented \cite{inv}. Among calculable models, the Nelson-Barr type
solution \cite{nb} has been studied later, chiefly because it can mimick
the Kobayashi-Maskawa CP violation at low energy. The spontaneous
CP violation occurs near GUT scale, which is different from
the earlier versions of calculable models using the
electroweak scale as the spontaneous CP violation scale. 

In 1985, the superstring axion was discovered, which is one of
the most important theoretical reasons for the axion to exist. 

\section{SUPERSTRING AXION}

If axion is the solution of the strong CP problem, it is better
to be present in string models. Indeed, axion is present
in any string models, but its relevance for the strong CP
solution depends on the compactification scheme.
Ten dimensional string models contain among massless spectrum the
bosons $G_{MN}$ ($MN$ symmetric), $B_{MN}$ ($MN$ antisymmetric), 
and $\phi$, where $M,N$ run through indices 
0,1,$\cdots$,9. Our interest here
is the antisymmetric tensor field $B_{MN}$ which contains
two kinds of axions: model-independent axion (MIa) \cite{witten} and 
model-dependent axions \cite{witten1}. 

The MIa is basically $B_{\mu\nu}$ where $\mu,\nu$ is the
4D indices $0,1,2,3$. The dual of the field strength is
defined as the derivative of MIa $a$,
\begin{equation} 
\partial^\sigma a\sim \epsilon^{\mu\nu\rho\sigma}H_{\mu\nu\rho},
H_{\mu\nu\rho}\sim \epsilon_{\mu\nu\rho\sigma}\partial^\sigma a
\end{equation}
The question is why we interpret this as an axion. It
is due to Green, Schwarz, and Witten \cite{gs,witten}.
The gauge invariant field strength $H$ of $B$ is 
\footnote{Here, we use the differential forms.} 
$H=dB-\omega^0_{3Y}+
\omega^0_{3L}$ with the Yang-Mills Chern-Simmons term
$\omega^0_{3Y}={\rm tr}(AF-A^3/3)$ and the Lorentz 
Chern-Simmons term $\omega^0_{3L}={\rm tr}(\omega R-\omega^3/3)$.
These satisfy $d\omega^0_{3Y}={\rm tr}F^2$ and $d\omega^0_{3L}
={\rm tr}R^2$. Therefore,
\begin{equation}
dH=-{\rm tr}F^2+{\rm tr}R^2
\end{equation}
Note also that one had to introduce a nontrivial gauge
transformation property of $B$. Then gauge anomaly is
completely cancelled by introducing the so-called
Green-Schwarz term \cite{gs}, 
\begin{equation}
S_{GS}\propto \int (B{\rm tr}F^4+\cdots)
\end{equation}
which contains the coupling of the form
\begin{eqnarray}
&\epsilon_{ijKLMNOPQR}B_{ij}F^{KL}F^{MN}F^{OP}F^{QR}\sim\nonumber\\
&B_{ij}(\epsilon_{\mu\nu\rho\sigma}F^{\mu\nu}F^{\rho\sigma})
<F_{kl}><F_{pq}>\epsilon_{ijklpq}\nonumber
\end{eqnarray}
where we note the Minkowski indices $\mu,\nu,\cdots$ and the
internal space indices $i,j,\cdots$. The nonvanishing VEV
$<F>$ gives $a'F\tilde F$ coupling at tree level. Thus we
are tempted to interpret $a'$ as an axion, and it was called
model-dependent axion \cite{witten1,kiwoon}. But we have to
check that there is no dangerous potential term involving
$a'$. But it has been shown that world-sheet instanton 
contribution
$$
i\int_{\Sigma_J}d^2z B_I\omega^I_{i\bar j}(\partial X^i\bar
\partial X^{\bar j}-\bar\partial X^i\partial X^{\bar j})
=2\alpha' B_J
$$
gives $a'(=B_I)$ dependent superpotential \cite{wen},
removing $a'$ as a useful degree for relaxing a
vacuum angle. In the above equation, $\alpha'$
is the string tension and $\omega^I_{i\bar j}$
represents the topology of the internal space,
\begin{equation}
B=B_{\mu\nu}dx^\mu dx^\nu+B_I\omega^I_{i\bar j}
dz^id\bar z^{\bar j}
\end{equation}
But $B_{\mu\nu}$ is still good since it does not get
a contribution from the stringy world-sheet instanton
effect. Eq. (33) implies
\begin{equation}
\Box a=-{1\over M}({\rm Tr}F_{\mu\nu}\tilde
F^{\mu\nu}-{\rm Tr}R_{\mu\nu}\tilde R^{\mu\nu})
\end{equation}
implying an effective Lagrangian of the form
\begin{equation}
{\cal L}={1\over 2}(\partial_\mu a)^2-{a\over 16\pi^2F_a}({\rm Tr}
F_{\mu\nu}\tilde F^{\mu\nu}-\cdots)
\end{equation}
Thus $a$ is the axion (MIa). Any string models have this
MIa and its decay constant is of order
$10^{16}$ GeV \cite{fmia}. At this point, we comment that there are two
serious problems of the superstring axion:

\noindent {\it (A) The axion decay constant problem}--It is known
that the decay constant of MIa is of order $10^{16}$ GeV
\cite{fmia} which is far above the cosmological upper bound 
of $F_a$ \cite{turner}. A large $F_a$ can be reconciled with
cosmological energy density if a sufficient number of radiation 
are added below 1 GeV of the universe temperature \cite{rad},
but then it is hopeless to detect the cosmic axion by cavity 
detectors. Therefore, $F_a$ is better to be lowered to
around $10^{12}$ GeV.  

\noindent {\it (B) The hidden sector problem}--It is a
general belief here that a hidden sector confining force,
e.g. $SU(N)$, is needed for supersymmetry breaking at
$10^{12}\sim 10^{13}$ GeV. If so, MIa gets mass also due
to the $a_{MI} F'\tilde F'$ coupling where $F'$ is the
field strength of the hidden sector confining gauge field,
and we expect $m_a\simeq \Lambda_h^2/F_a$ which is obviously
too large to bring down MIa to low energy scale for the
solution of the strong CP problem. For example, the axion
gets potential both from the hidden sector scale $\Lambda_h$
and the QCD scale $\Lambda_{QCD}$ (if there is no matter)
in the follwing way,
$$
-\Lambda_{QCD}^4\cos({a_{MI}\over F_a}+\theta^0)-\Lambda_h^4
\cos({a_{MI}\over F_a}+\theta_h^0)
$$
where we added two terms with independent phases $\theta^0,
\theta_h^0$ which arise at the string scale when CP is
broken. Because $\Lambda_h\gg\Lambda_{QCD}$, the vacuum
chooses ${a_{MI}\over F_a}+\theta_h^0=0$, i.e. $<a_{MI}>\simeq
-\theta_h^0 F_a$, implying $\bar\theta\simeq\theta^0-\theta_h^0$
which is not zero in general. If we want to settle both $\theta_h$
and $\bar\theta$ at zero, then we need two independent axions.
However, it is known that only MIa is available at string induced 
low energy physics. Therefore, we say that there is the hidden
sector problem in the MIa phenomenology.

The above two problems have to be resolved if the string theory
render an acceptable low energy standard model and also if
the axion solution has a profound root in the fundamental theory
of the universe. It turns out that it is very difficult to
achieve. Only in a limited case, it may be possible to
find a possible route.

One such example is the so-called anomalous $U(1)$ models
\cite{anom} in which the gauge group takes the form
$U(1)_A\times SU(3)\times SU(2)\times U(1)\times\cdots$.
The gauge group $U(1)_A$ has an anomaly if one considers
matter fields only, but in the whole theory there is no
anomaly. The Green-Schwarz term contains
\begin{eqnarray}
&\epsilon_{MNOPQRSTUV}B^{MN}\cdot {\rm Tr}F^{OP}\nonumber\\
&\cdot <F^{QR}><F^{ST}><F^{UV}>
\end{eqnarray}
which introduces the coupling $M_c(\partial^\mu a_{MI})A_\mu$.
Namely, the MIa becomes the longitudinal degree of freedom of
$A_\mu$. Thus, the $U(1)_A$ gauge boson becomes massive
and $a_{MI}$ is removed at low energy. Below the scale
$M_c$, then there exists a global symmetry \cite{ami,gkn}. 
Superstring models also need an extra
hidden sector confining force. Then, even if
the MIa is present, it obtains a dominant contribution to the mass 
from the hidden sector instanton effects. One can make the
contribution to the MIa mass absent if there is a massless
hidden colored fermion. The first choice is the theory
of a massless hidden sector gaugino without a hidden matter.
But the hidden sector gaugino is NOT massless. Nevertheless,
the final hidden sector gaugino mass is not introduced by hand, but
it arises from the condensation of the hidden sector gauginos,
thus the contribution to the MIa potential is absent in this
limit. However, the string (or gravitational) theory does not
preserve any global symmetry, thus there must be interactions
violating the $R$ symmetry.\footnote{With the massless gaugino
with only renormalizable gaugino self interactions, there is an
$R$ symmetry.} But the contribution to the potential from
the $R$ violating terms is much smaller than a naive
dimensional counting. In addition, if there survives a
discrete subgroup of $R$ symmetry, then the contribution to
the MIa potential is sufficiently suppressed \cite{gkn}.
Thus, there is a hope of introducing a very light axion
in superstring models.

\section{CONCLUSION}

I reviewed the very light axions from the particle theory
viewpoint. The strong CP problem is real in QCD,\footnote{
There exists a comment that it is not a real problem \cite{okubo},
but there exists other arguments claiming that it is real.} 
and as we have seen the axion solution is the most elegant one.
If it exist, it can contribute to the astrophysical
and cosmological evolution. If the parameters of the theory
is in the right range, it can be detected, which is the
reason that this workshop is so interesting. Theoretically,
on the other hand, the superstring is expected to give
the low energy standard model. Therefore, if the very light
axion is to solve the strong CP problem, it is better to be
implemented in the superstring models. If the superstring
derived standard model is found and it contains a very light
axion, one can calculate $c_{a\gamma\gamma}$ with certainty.
On the contrary if the very light axion is found, then it
can lead to discovering the superstring standard model.

\end{document}